\documentclass[usegraphicx]{mn2e}

\def\Msun{\ifmmode{~M_\odot}\else$M_\odot$~\fi}
\def\Mden{\ifmmode{~M_\odot $pc$^{-3}}\else$M_\odot $pc$^{-3}$\fi}
\def\kms{\ifmmode{$~km\thinspace s$^{-1}}\else km\thinspace s$^{-1}$\fi}

\def\Msun{\ifmmode{~{\rm M}_\odot}\else${\rm M}_\odot$~\fi}
\def\kms{\ifmmode{$~km\thinspace s$^{-1}}\else km\thinspace s$^{-1}$\fi}
\def\etal{\ifmmode{${\it et al.}$}\else {\it et al.}\fi}
\def\Mden{\ifmmode{~{\rm M}_\odot~{\rm pc}^{-3}}
 \else${\rm M}_\odot$~pc$^{-3}$\fi}
\def\la{\mathrel{\mathchoice {\vcenter{\offinterlineskip\halign{\hfil
$\displaystyle##$\hfil\cr<\cr\noalign{\vskip1.5pt}\sim\cr}}}
  {\vcenter{\offinterlineskip\halign{\hfil$\textstyle##$\hfil\cr<\cr
  \noalign{\vskip1.0pt}\sim\cr}}}
  {\vcenter{\offinterlineskip\halign{\hfil$\scriptstyle##$\hfil\cr<\cr
  \noalign{\vskip0.5pt}\sim\cr}}}
  {\vcenter{\offinterlineskip\halign{\hfil$\scriptscriptstyle##$\hfil
  \cr<\cr\noalign{\vskip0.5pt}\sim\cr}}}}}
  \def\ga{\mathrel{\mathchoice {\vcenter{\offinterlineskip\halign{\hfil
$\displaystyle##$\hfil\cr>\cr\noalign{\vskip1.5pt}\sim\cr}}}
  {\vcenter{\offinterlineskip\halign{\hfil$\textstyle##$\hfil\cr>\cr
  \noalign{\vskip1.0pt}\sim\cr}}}
  {\vcenter{\offinterlineskip\halign{\hfil$\scriptstyle##$\hfil\cr>\cr
  \noalign{\vskip0.5pt}\sim\cr}}}
  {\vcenter{\offinterlineskip\halign{\hfil$\scriptscriptstyle##$\hfil
  \cr>\cr\noalign{\vskip0.5pt}\sim\cr}}}}}

\def\bea{\begin{array}}
\def\eea{\end{array}}
\def\beq{\begin{equation}}%
\def\eeq{\end{equation}}
\def\ben{\begin{eqnarray}}
\def\een{\end{eqnarray}}

\def\sech{\mathop{\rm sech}\nolimits}

\def\spose#1{\hbox to 0pt{#1\hss}}
\def\lta{\mathrel{\spose{\lower 3pt\hbox{$\mathchar"218$}}
     \raise 2.0pt\hbox{$\mathchar"13C$}}}
\def\gta{\mathrel{\spose{\lower 3pt\hbox{$\mathchar"218$}}
     \raise 2.0pt\hbox{$\mathchar"13E$}}}

\def\kms{\mbox{$\,{\rm km}\,{\rm s}^{-1}$}}
\def\kpc{\mbox{$\,{\rm kpc}$}}
\def\pc{\mbox{$\,{\rm pc}$}}
\def\parc{\mbox{$\,{\rm pc}$}}
\def\Msun{\mbox{$\,{\rm M}_\odot$}}

\begin{document}
\title{Simulations of the heating of the Galactic stellar disc}

\author[J. H\"anninen and C. Flynn] {Jyrki H\"anninen$^1$ and Chris
Flynn$^{1,2}$ \\ $^1$Tuorla Observatory, V\"ais\"al\"antie 20, FIN-21500,
Piikki\"o, Finland\\ $^2$Centre for Astrophysics and Supercomputing, Swinburne
University of Technology, Hawthorn, Australia}


\maketitle

\begin{abstract}

The velocity dispersion of nearby stars in the Galactic disc are well
known to increase substantially with age; this is the so-called
Age-Velocity relation, and is interpreted as a ``heating'' of the disc
as a function of time.  We have studied the heating of the Galactic
stellar disc due to giant molecular clouds and halo black holes, via
simulations of the orbits of tracer stars embedded in a patch of the
local Galactic disc. We examine a range of masses and number densities
of the giant molecular cloud and halo black hole perturbers. The
heating of the stellar disc in the simulations is fit with a simple
power law of the $\sigma \propto t^\alpha$ where $\sigma$ is the
velocity dispersion of the tracer stars as a function of time, $t$. We
also fit this form to the best determinations of the increase in the
velocity dispersion as a function of time as derived from stars in the
solar neighbourhood for which ages can be reliably
assigned. Observationally, $\alpha$ is found to lie in the range 0.3
to 0.6, i.e. it remains poorly constrained and its determination is
probably still dominated by systematic errors. Better constrained
observationally is the ratio of the velocity dispersions of the stars
in the vertical $z$ and horizontal $x$ (i.e. toward the Galactic
center) directions, being $\sigma_z/\sigma_x = 0.5 \pm 0.1$.

For the heating of the stellar disc due to giant molecular clouds (GMCs) we
derive a heating $\sigma \propto t^{0.21}$, which differs somewhat from early
(analytic) studies in which $\sigma \propto t^{0.25}$. This confirms the well
known results that there are insufficient GMCs to heat the Galactic disc
appropriately. A range of dark halo black hole scenarios are verified to heat
the stellar disc as $\sigma \propto t^{0.5}$ (as expected from analytical
studies), and give $\sigma_z/\sigma_x$ in the range 0.5 to 0.6, which is
consistent with observations. Black holes with a mass of $10^7$ \Msun\ are our
favoured disc heaters, although they are only marginally consistent with
observations. Simulations featuring a combination of giant molecular clouds and
halo black holes can explain the observed heating of the stellar disc, but
since other perturbing mechanisms, such as spiral arms, are yet to be included,
we regard this solution as ad hoc.

\end{abstract}


\section{Introduction}

The random motions of stars near the Sun are well known to increase with
stellar age, and this effect is known as the disc Age Velocity Relation
(hereafter AVR). Broadly speaking, the total velocity dispersion of stars rises
from $\approx $20 \kms\ at the lowest measurable ages to $\approx$ 60-80 \kms\ at
an age of $\approx$ 10-12 Gyr. The ``heating'' of the orbits is initially rapid
but levels off after some Gyr. The heating is not uniform in each of the three
velocity components; the velocity dispersion rises to a higher value in the
Galactic plane than in the vertical direction.

A number of mechanisms have been proposed to explain the AVR, in particular the
gravitational perturbative effect on stellar orbits by giant molecular clouds
(GMCs) in the Galaxy's gas layer \cite{spit51,spit53}, by spiral arms in the
disc \cite{bawo}, or by massive black holes in the Galactic dark halo (BHs)
\cite{lacey85}. Dramatic disc heating also takes place when satellite galaxies
fall onto galactic discs \cite{quinn93}, and such events may well leave an
abrupt feature in the AVR, but if the rain of such satellites is relatively
gentle they might also heat discs smoothly \cite{vel99}.

Spitzer and Schwarzschild \shortcite{spit51,spit53} first argued that the
observed growth of the stellar velocity dispersion could be explained by
encounters of disc stars with GMCs with masses of $M_{\rm GMC} \sim 10^{6}
\Msun$. The main difficulty with this scenario is that the observed number of
GMCs modify stellar orbits in a manner inconsistent with observations.
Firstly, they are too few to reproduce the observed heating, and secondly, too
much heating is produced in the vertical direction relative to the disc plane
\cite{lacey84}. However, GMCs do exist so they certainly provide a mechanism
for redirecting orbits out of the Galactic plane, whatever is responsible for
the heating. Heating due to transient spiral arms suffers the opposite problem
that the amount of vertical heating is too low compared to observations
\cite{carl87}.

This naturally leads to the idea that the observed heating could be explained
by the combined effect of the transient spiral arms and giant molecular
clouds. In this model spiral density waves would act as main source of heating
and GMCs would delect the stellar orbits out of the Galactic plane, thus
creating the required velocity dispersion in the vertical direction
\cite{carl87}.  However, these calculations are rather complicated: e.g. the
relative effectiveness of spiral and GMC heating has to modelled with an
empirical parameter. The rate at which spiral features heat a stellar disc
depends on the potential's spatial pattern and on the time variability of the
pattern. In principle, spatial stuctures of galaxies can be determined from
photometry of real galaxies and temporal information about spiral stucture can
be obtained from numerical simulation and dynamical theory. In practice,
however, we do not have adequate information about strengths, growth rates or
duty cycles of spiral features. So empirical parametrization has so far been
the only possible approach \cite{jebi,je}.

Lacey and Ostriker \shortcite{lacey85} proposed that the Galactic dark halo
might consist of massive black holes (BHs) which would heat the disc as they
pass through it. The total heating produced by their proposed $2 \times 10^6$
\Msun\ black holes is consistent with observations, but the velocity ellipsoid
was found to be quite round, which is inconsistent with observations.
Furthermore, if the dark halos of dwarf spiral galaxies were dominated by
similar BHs their discs would be easily destroyed \cite{friese95}. Finally, BHs
passing through the disc might be expected to accrete and reveal themselves as
high proper motion X-ray sources, and no such objects have been seen. It seems
worth noting though that the central black hole in the Milky Way demonstrates
that the accretion rate and emergent flux from a black hole of about the
desired mass ($2.6 \times 10^6$ \Msun) is far from well understood
\cite{genz98}; the central black hole in the Milky Way is remarkably quiet in
X-rays. Another way of detecting these putative black holes is by statistically
studying large numbers of orbits of nearby stars; this may be possible with
very accurate distances and kinematics which will become available for stars
within some kiloparsecs from ESA's GAIA satellite.

In order to avoid problems related to the black hole scenario Carr and Lacey
\shortcite{carr87} proposed dark clusters of less massive objects. The latest
generation of cosmological simulations of galaxy formation do indeed show quite
clumpy dark halos around Milky Way type spirals \cite{moore99}. The amount,
size and distribution of these clumps is still very uncertain, mainly due to
numerical resolution issues, but if they can survive in the inner regions of
dark halos they may well be able to heat discs in a manner consistent with
observations.

With this in mind we have embarked on a new study of disc heating simulations,
starting with examining heating due to GMCs and BHs. The European Space
Agency's Hipparcos satellite has provided a wealth of new data on the distances
and kinematics of nearby stars, so that one can now construct very much
improved measurements of ages and kinematics. The velocities of the stars are
much improved through the excellent Hipparcos parallaxes, while the ages of
stars can be much better estimated, particularly for stars near the disc main
sequence turn-off, because of the greatly improved absolute magnitudes.

Ideally, one would like to carry out a self-consistent N-body simulation of an
entire disc galaxy, with and without the perturbing influences due to GMCs/BHs
etc, but this is not yet possible even with the fastest computers. The
computation would involve both the stars (at approx 1 \Msun) and the perturbers
(for a range of masses, $10^5$-$10^7$ \Msun), implying some $10^{11}$ stars and
thousands of GMCs to millions of BHs in the simulations. This is not yet
feasible in an N-body simulation. In this paper we render the problem tractable
by simulating the heating in a local patch of the Galactic disc consisting
typically of few tens of thousands of (massless) tracer stars, embedded within
a fixed Galactic background potential, through which massive perturbers
move. By using a local simulation method we can resolve very well the effects
of perturbers on nearby stellar orbits, for comparison with the locally
obtained observations. A very similar method has been used by Fuchs et
al. \shortcite{fuchs94} who studied disc heating due to GMCs and/or $5 \times
10^6 \Msun$ black holes. We are able to broadly reproduce the results of their
code in this area, but we also explore a larger range of scenarios.

From our simulations, we find that GMC heating is less effective than earlier
thought and further that heating due to GMCs creates a flatter velocity
ellipsoid than found in earlier studies. We find that BHs are not as effective
in heating the disc as earlier thought. However, very massive halo black holes
($M_{BH} = 1 \times 10^7 \Msun$) can heat the disc up to the observed amount,
but the results of the heating rate and the ratio of the velocity dispersions
are only marginally consistent with observations.  A well selected combination
of BHs and GMCs can give the observed heating in every component of the
velocity dispersion within the observational error limits, but while
interesting this solution is probably ad hoc, as the simulations do not yet
model the effects of other perturbation sources, such as spiral waves. This
will be examined in future work.

This paper is organised as follows. In section 2 we review pre- and
post-Hipparcos work on the observational determination of the age-velocity
relation (AVR) for the local disc. We find the AVR is much less well measured
than has previously been assumed. Observational issues thus still fundamentally
limit our ability to discriminate between disc heating mechanisms.  In section
3 we describe briefly our simulation method and numerical accuracy. In sections
4, 5, and 6 we demonstrate how massive perturbes (GMCs and halo BHs) heat up
the stellar disc. We draw our conclusions in section 7.

\section{Observations}

The Hipparcos satellite measured parallaxes and space motions for a complete
sky coverage of about 60,000 stars, with a further 60,000 stars assembled from
a variety of pre-existing sources. As a result, the kinematics of stars used to
construct the relationship between mean stellar age and the components of the
velocity dispersion (the Age-Velocity relation, AVR) have been greatly
improved. We briefly review here the state of the observational AVR both
pre-Hipparcos and post-Hipparcos.

\subsection{Observational Age-Velocity-Relation} 

Figure 1(a), shows several determinations of the AVR from the literature. We
divide the measurements into pre-Hipparcos and post-Hipparcos determinations.

Each AVR presented and the simulation results has been fit with the following
form

\begin{equation}
\label{fit}
\sigma(t) = \sigma_\circ (1 + \frac{t}{\tau})^\alpha \ 
\end{equation}

where $\sigma(t)$ is the total velocity dispersion as a function of time, $t$,
$\sigma_\circ$ is the initial velocity dispersion at $t = 0$, $\tau$ is a
constant with unit of time, and $\alpha$ is the heating index
\cite{wie,villu85}. Our motivation here is to fit the data with a simple
analytical law which can also be fit to the simulations.

\subsubsection{Pre-Hipparcos determinations}

Firstly, consider the pre-Hipparcos measurements of the AVR, which are shown in
panel $a_1$ of Figure 1.  Edvardsson et al's \shortcite{edva93} data set of
accurately determined ages for 189 F and G stars is shown by crosses.  Wielen's
\shortcite{wie} determination, based mainly on K and M dwarfs which have been
ranked by age based on chromospheric emission line measurements, is shown by
squares. An improved version of the data set \cite{fuchs00} is shown by
triangles. These latter data sets are mostly M dwarf types, for which the
kinematical and parallax measurements have only been marginally improved by
Hipparcos, and they remain robust in the post-Hipparcos epoch.

We have fit each data set to Eqn \ref{fit}, using a non-linear least squares
gradient-expansion algorithm (using the IDL software package).  All data points
were given equal weight. For these three data sets, we find heating indices
$\alpha$ in the range 0.47 to 0.61 (see Table 1).

\setcounter{table}{0}
\begin{table}
\begin{center}
\caption{Values of the heating index obtained by fitting data in the
literature.  The first five rows show our fits, and the last two show the
author's own reported values. Note that Holmberg (2001) removes the effect of
higher velocity thick disc stars when making his (more reliable) estimate of
the heating index, whereas we leave them in (in order to estimate the bias they
impose). There is a very considerable range in the derived values of $\alpha$,
which far exceeds the reported (internal) error estimates; $\alpha$ is
therefore very likely to be dominated by systematic errors in the methods used
to estimate stellar ages.}
\begin{tabular}{lc}
\hline
Study                    &     $\alpha$    \\
\hline
R. Wielen (1977)         &   $ 0.61  \pm  0.04$  \\
Fuchs et al. (2000)      &   $ 0.59  \pm  0.04$  \\ 
Edvardsson et al.(1993)  &   $ 0.47  \pm  0.002$ \\
H. Rocha-Pinto           &   $ 0.26  \pm  0.008$ \\
Holmberg (2001) All stars         &   $ 0.45  \pm  0.04$  \\ 
Holmberg (2001) Thick Disc removed         &   $ 0.33   \pm  0.03 $  \\
Binney et al. (2000)     &   $ 0.33   \pm  0.03  $ \\
\hline
\end{tabular}\label{atab}
\end{center}
\end{table}

Our motivation is to fit the same form of the heating law to all the available
data and to all the simulations.We should note that the heating indices
$\alpha$ we derive are not quite the same as those derived by the authors
above, who apply additional physical constraints to the fitting (e.g. the
nature of the stellar orbital diffusion, and the weight to be applied to the
initial value of the velocity dispersion). Nevertheless, the fitted indices are
similar.
 
\setcounter{figure}{0}
\begin{figure*}
\label{obsig}   
\input epsf
\center
\leavevmode
\epsfxsize=0.9
\columnwidth
\epsfbox{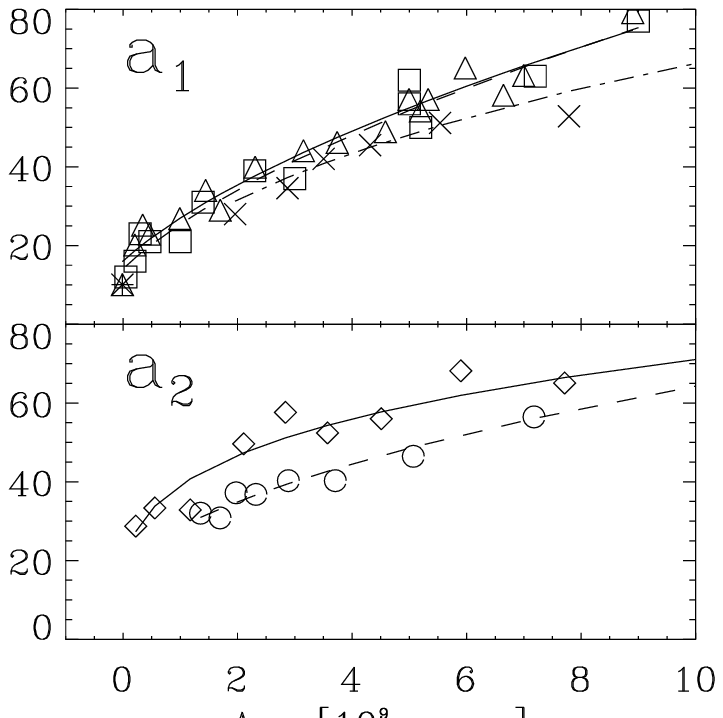}
\epsfbox{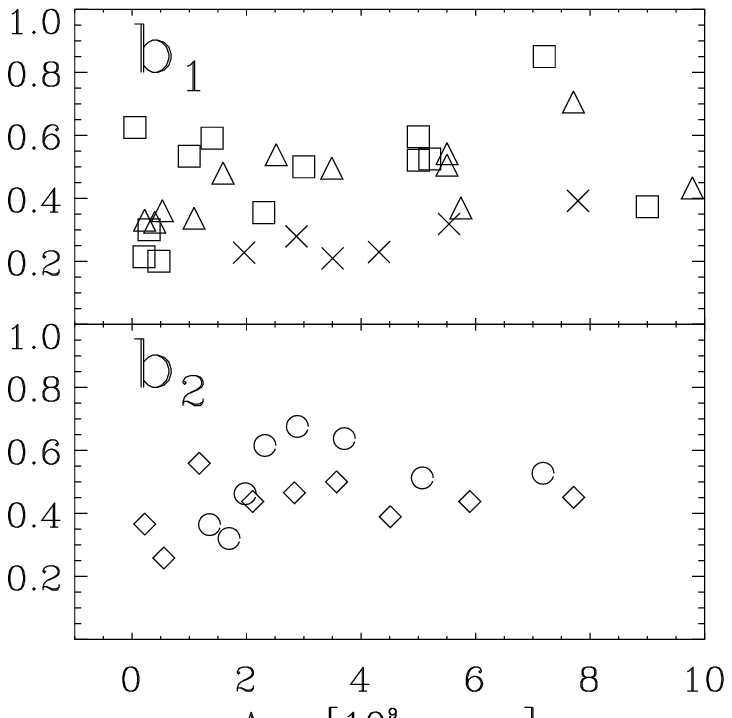}
\caption{The observed total stellar velocity dispersion and ratio of velocity
dispersions for nearby disc stars, as obtained by different authors. Best fits
to the heating laws as parameterised by Eqn. 1 are also shown. Panel ($a_1$):
the solid line is the fit to Fuchs et al. data (triangles), the dashed line is
the fit to Wielen's data (squares), and the dot-dashed line is the fit to
Edvarsson et al. data (crosses). Panel ($a_2$): the solid line is the fit to
Rocha-Pinto's data (diamonds) and the dashed line is the fit to Holmberg's data
(circles).  Panels ($b_1$) and ($b_2$) show the observed ratios of the vertical
to the radial velocity dispersion for the same data sets and using the same
symbols.}
\end{figure*}

\subsubsection{Post-Hipparcos determinations}

We now consider post-Hipparcos AVR determinations (Figure 1($a_2$)).  Circles
show data from Holmberg \shortcite{holm2001}, for 1486 Hipparcos F and G type
stars for which ages can be determined with an accuracy of about 30\% (of the
age), by fitting absolute magnitude and colours to stellar isochrones. We show
by diamonds our analysis of data kindly provided in advance of publication by
H. Rocha-Pinto, who has determined ages of 425 mainly F, G and K stars using a
carefully calibrated technique based on chromospheric emission \cite{rocha98}.
For his stars we have measured the total velocity dispersion as a function of
age for the objects with [Fe/H]$ > -0.5$, being careful to exclude several high
velocity outliers.

When fitting the data from H. Rocha-Pinto we yield a heating index
$\alpha = 0.26$ with a very small (formal) error (see Table
\ref{atab}). However, the value of the heating index is sensitive to
the fitting procedure.  For example, if we had used bins of 60 stars
instead of 40-star-bins, we would have obtained $\alpha = 0.29 \pm
0.02$.  Because most of the stars in the sample were young we binned
the data so that there were equal number of stars in each bin rather
than using bins of equal width in age.

The data from Holmberg was fitted in similar manner: the objects with [Fe/H]
$<-0.5$ were removed from the data set and equal number bin were used.  We
obtained $\alpha = 0.45 \pm 0.04$ for the heating index. This is close to
Holmberg's own results ($\alpha = 0.46 \pm 0.02$) for all his stars. However,
Holmberg extends his fitting procedure by separating the stellar data into thin
and thick disc components by a maximum likelihood method, deriving a best
fitting AVR for the thin disc for which the heating index is $\alpha = 0.33 \pm
0.03$.

Binney et al. \shortcite{bin2000} have used a different approach to constrain
the star forming history and velocity dispersion evolution in the Solar
neighbourhood. They have used photometrically complete sample of 11,865
Hipparcos stars. By combining local colour-magnitude diagram with the Padua
isochrones and, crucially, kinematic information, they have obtained a heating
index of $\alpha = 0.33 \pm 0.03$ and a disc age of $11.1 \pm 0.8$ Gyr.  The
value obtained is consistent with Holmberg's, for a smaller set of stars drawn
from the same source (Holmberg works with individually determined stellar
ages, and limits himself only to those for which the age can be determined
quite accurately).

While the values of $\alpha$ obtained by Binney et al. \shortcite{bin2000} and
Holmberg \shortcite{holm2001} are consistent, they are inconsistent with the
value we derive from Rocho-Pinto's data set, which uses an independent method
to estimate stellar age, and inconsistent with the heating index derived for
the K and M dwarfs, which have been age ranked via chromospheric emission
\cite{fuchs00}. Fuchs et al. \shortcite{fuchs00} show that the discrepency
between values for the heating index of 0.3 and 0.5 is actually rather mild
(see their Fig. 1), when plotted in the age-velocity dispersion plane.

The data sets give significantly different values for the heating index,
ranging from 0.3 to 0.5. The error in the ages does give rise to a small
systematic error in $\alpha$; we have performed Monte-Carlo simulations in
which we change the ages of the observed stars by a Gaussian distributed random
fractional error, and recomputed the fit for $\alpha$. Random age errors of
order 20\% typically increase $\alpha$ slightly, typically by less than 0.05. A
20\% error in stellar ages is realistic in the context of determining the
heating index in the AVR. For example, Holmberg uses only stars with a
fractional error of typically 20-30\% (and always less than 50\%), while
Rocha-Pinto (private communication) reports fractional errors of order 30\%.
However, for the youngest stars in the sample, the errors may be significantly
larger, of order 0.5 in log(age). Monte-Carlo simulations show that this is
certainly enough to reduce the heating index from 0.5 to 0.3. Furthermore, the
determination of $\alpha$ is sensitive to systematic errors in the ages of the
younger stars. We have performed Monte-Carlo simulations which show that
changes in the value of $\alpha$ of order 0.1 can be obtained if, for example,
there is a systematic error of 50\% in the assigned ages of stars younger than
a few Gyr (as might occur as a result of systematic errors in isochrone
colours). There are thus several mechanisms which could produce systematic
errors in the heating index as derived from observations.

We are reluctant to arbitrarily assign any of these data sets greater credence
than any other; they are all very carefully constructed and they use different
techniques, some independent, to assign ages to the stars. Instead we move on
to consider the ratio of the velocity dispersions, rather than the vertical
velocity dispersion alone. This quantity turns out to be much less sensitive to
the ages of the stars and turns out to be very useful in comparing the data to
the simulations.

\subsection{Ratio of Velocity Dispersions} 

In panels ($b_1$) and ($b_2$) of Figure 1 the ratio of the vertical to the
radial velocity dispersion $\sigma_z / \sigma_x$ is plotted for the same cases
as in panels ($a_1$) and ($a_2$) and using te same symbols. The scatter seems
to be larger amongst the pre-Hipparcos observations (panel ($b_1$)) than in the
post-Hipparcos data (panel ($b_2$)). The latter observations indicate that the
ratio is $\sigma_z / \sigma_x \simeq 0.5$ and the former observations
corroborate this. Using the Hipparcos data, Dehnen and Binney \shortcite{debi}
have determined the velocity dispersion evolution as a function of colour index
$B-V$. Their results also indicate that the ratio $\sigma_z / \sigma_x$ tends
to $\sim 0.5$.

\subsection{Comparing observations and simulations} 

The observations show that the heating index lies between $\alpha = 0.3$ and
$0.5$: in other words $\alpha$ is not yet well constrained by the
observations. This was a major uncertainty in comparing the results of the
simulations and observations. On the other hand, the ratio $\sigma_z /
\sigma_x$ is quite well measured at $0.5 \pm 0.1$. The ratio of vertical
velocity dispersion to radial velocity dispersion thus turned out to be the
best discriminator when theoretical models or numerical simulation are compared
with observations (Sections 4, 5, and 6).

\section{Simulation method}

The local simulation method was first used in planetary ring dynamics by Wisdom
and Tremaine \shortcite{wite} (see also Salo 1995) and was first applied to
disc galaxies by Toomre \shortcite{toom}.  In the method all calculations are
restricted to a small co-moving box within the Galactic disc. A typical
computer run simulates a ``patch'' of the disc, being $1\times 1$ or $2\times
2$ kpc square, and 2 to 4 kpc in vertical extent, containing of order $10^4$
tracer stars in the disc as well as any perturbers (GMCs or BHs).

The coordinates of stars in the region are referred to a point which moves on a
circular orbit around the Galactic center at a distance $r$. A rotating
Cartesian coordinate system is placed at the reference position, the $x$-axis
pointing radially outward (away from the Galactic center), the $y$-axis in the
direction of orbital motion, and the $z$-axis in the direction normal to the
Galactic plane. The orbits of stars in the patch are computed by integrating
the linearized equations of motion \cite{hill78,juto66}:

\begin{eqnarray}
\label{Hill}
\ddot{x} - 2 \Omega \dot{y} + (\kappa^2 - 4 \Omega^2) x &= F_x \cr 
\ddot{y} + 2 \Omega \dot{x} &= F_y \cr
\ddot{z} + \nu^2 z &= F_z 
\end{eqnarray}

where $\Omega$ is the orbital frequency, and $\kappa$ and $\nu$ are the
epicyclic frequency and the vertical frequency of motion, respectively.  The
numerical integrator is the RK4 routine \cite{press}. We adopt Solar
neighbourhood ($r=8$ $\kpc$) values of $\Omega = 25.9 \kms /\kpc$, $\kappa =
36.0 \kms /\kpc$, and $\nu = 98.7 \kms /\kpc$ \cite{bitre}. The tracer stars
feel the Galactic potential and the forces due to perturbers (GMCs or black
holes).  In order to avoid numerical artefacts, the gravitational forces ($F_x,
F_y, F_z$) are calculated so that Newton's third law is satisfied, i.e. a
perturber also feels the gravitational force from tracer stars.

\subsection{Numerical accuracy}

All the calculations are performed in double precision. In an unperturbed test
run with an integration step size $\Delta t = T_{\mathrm orb}/800$ the relative
error in the stellar vertical kinetic energy was observed to be $\vert \Delta
E_z / E_z(0) \vert < 4 \times 10^{-7}$ at the end of the simulation run
($t_{\mathrm END} = 50 T_{\mathrm orb} \simeq$ 12 Gyr). Here $T_{\mathrm orb}
\simeq 237.23 \times 10^6$ years, is the orbital time around the Galactic
center.  During the run the vertical component of the system angular momentum
relative to the reference position should remain zero, and is conserved to
better than $\vert \Delta I_z \vert / \sqrt{\overline{I_z^2(0)}} < 3 \times
10^{-14}$ in which we have compared the change in the system angular momentum
to the mean initial angular momentum of the stars.

In the production runs we have used a still shorter integration step
size: from $\Delta t = T_{\mathrm orb}/1000$ to $\Delta t = T_{\mathrm
orb}/16000$ depending on the size and mass of the perturber. We have
used such a short time step because the numerical accuracy of the
orbits is dominated by the close encounters with the massive
perturbers. Most of the numerical error is due to these instances. The
time step would be even shorter if we were interested in the orbits,
but it is actually the statistical heating of many orbits which
concerns us, rather than maximum accuracy for particular orbits.  The
time step was determined by simply testing how a shorter and shorter
time step affects the evolution of the velocity dispersion. Because
the integration error creates artificial noise in the system, we just
need to search for the largest step size for which the velocity
dispersion of the tracer stars does not increase due to step size in a
simulation with massive perturbers. The optimal step size will
naturally be different for a point mass like black holes and for an
extended object like a GMC.

An important potential source for systematic error is in the calculation of
gravitational force. In practice, because we are using a local simulation box,
the calculation of the gravitational interaction has to be cut off at some
distance. Furthermore, in a local simulation method this limit should be at
most half of the box size, in order to avoid computing the gravitational force
from a perturber more than once.  We need to choose a box size for which the
heating of the orbits by distant encounters outside the cut-off-radius is
insignificant compared to closer encounters within the cut-off-radius.  We have
performed sets of simulations in order to test for the appropriate patch
size. Some examples will illustrate this: for a $1 \times 1 \kpc^2$ patch and
black holes with $M_{BH} = 1 \times 10^{6} \Msun$ we find the total stellar
velocity dispersion after 50 Solar orbital periods rises to $\sigma_{tot} =
22.7$ \kms; changing to a patch of $2 \times 2 \kpc^2$, and running the
simulation again with a larger cutoff length for gravity, we derive a total
velocity dispersion of $22.1$ \kms\ at the end of the run. The difference
between the runs is within the simulation noise level and we conclude that the
effects of long range forces are simulated with good accuracy. For GMCs of
similar mass the patch size $1 \times 1 \kpc^2$ was observed to be too small,
and minimum size $2 \times 2 \kpc^2$ had to be used.

\subsection{Boundary conditions and the tracer stars}

Tracer stars move on epicycles in the patch, and can potentially cross the
patch's boundaries. Periodic boundary conditions are used to recover these
stars, which enter from the opposite boundary with a correction made for
Galactic shear, as described in detail by Wisdom and Tremaine
\shortcite{wite}. The box in which the calculations are performed is thus
effectively surrounded by virtual boxes which shear due to Galactic rotation,
as illustrated in Figure 2.  We typically use boxes which are $1 \times 1$ kpc
or $2 \times 2$ kpc in the Galactic plane, and extend 2 or 4 kpc above and
below it.

The linearized equations are valid as long as $\vert x \vert$ and $\vert z
\vert \ll r$. For all our simulations the velocity dispersions remain low and
this condition is met.

\setcounter{figure}{1}
\begin{figure}
\label{boxfig}   
\input epsf
\center
\leavevmode
\epsfxsize=0.9
\columnwidth
\epsfbox{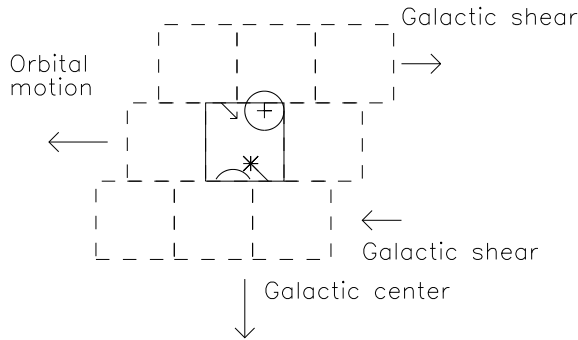}
\caption{The simulation box (solid line) and its surrounding virtual boxes
(dashed lines) are illustrated. Gravitational forces on a given star (cross)
are calculated from the gravitational perturbers whose nearest image lies
within the distance $R_{g}$ (denoted by the circle). If a star (asterisk)
leaves the simulation cell, its image will enter the cell in a way defined by
the boundary conditions.}
\end{figure}

\subsection{Massive Perturbers} 

We simulate two types of massive perturber: Giant Molecular Clouds (GMCs) and
Black Holes (BHs). The GMCs are placed in the Galactic plane on near-circular
orbits, which are then integrated in the same manner as the tracer stars (but
without the effect of the other perturbers, as this would unnecessarily slow
the simulation and heat up the GMC population). The same periodic boundary
conditions which replaces those stars which leave the box are also applied to
the GMCs.

The BHs are on orbits characteristic of the dark halo, with high velocities
relative to the disc stars. BHs move through the box on essentially straight
line orbits.  They obey the same periodic boundary conditions as the tracer
stars in the radial ($x$-) and azimuthal ($y$-) direction, but not in the
vertical ($z$-) direction.  The border in the $z$-direction is given special
status, by setting it to be far enough from the disc midplane, in order that no
forces need to be calculated across it. In practice one must take care that the
box is large enough in the vertical direction, so that the stars never get
close to the vertical boundary. When a halo black hole goes through the
vertical border, it is removed from the simulation and a new one is randomly
created from the parent velocity and space distribution.  The number density of
dark halo BHs entering and leaving the box is maintained at a constant level.

The orbits of the tracer stars are computed by solving Equation \ref{Hill} in
the presence of massive perturbers in the disc or moving through the disc from
the Galactic halo, by direct summation. The forces are calculated on each
tracer star for all perturbers which are within a radius $R_{g}$ of the star
(see Fig. 2). In practice we have always used the maximum distance allowed by
the patch, {\em i.e.} the calculation radius is half of the box width.

\section{Disc Heating by Giant Molecular Clouds}

\setcounter{figure}{2}
\begin{figure*}
\label{fig2a}   
\input epsf \center \leavevmode \epsfxsize=0.9 \columnwidth
\epsfbox{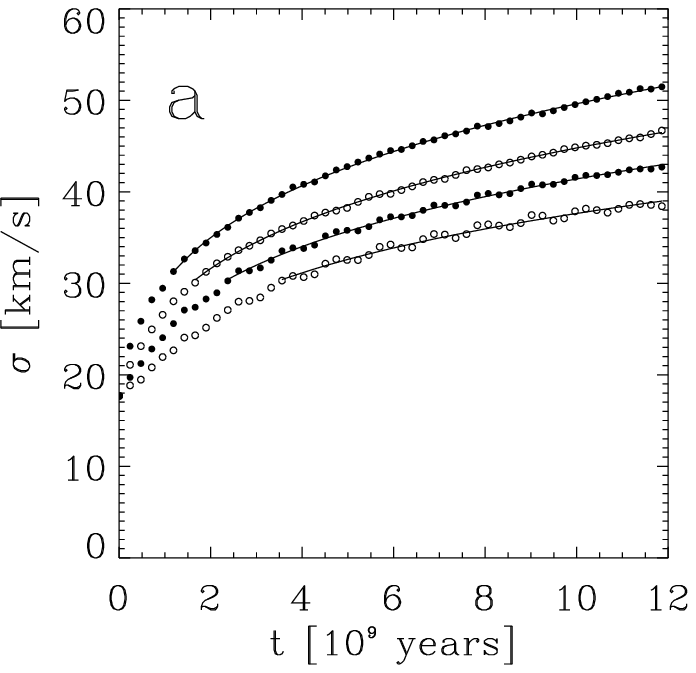}
\epsfbox{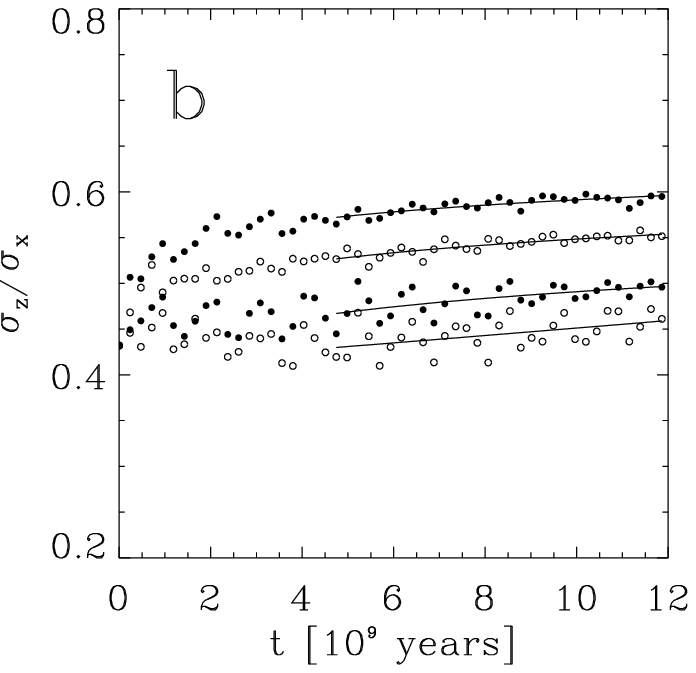}
\caption{Panel (a). The evolution of total velocity dispersion $\sigma$ due to
GMCs. Symbols are the simulation results and the curves represent least squares
fits of the Eq. \ref{fit} to the points.  Panel (b): The ratio of the vertical
to the radial velocity dispersion shown as a function of the number density of
GMCs, $\Sigma_{\rm GMC}$, in the simulation box. In both panels, from bottom to top,
$\Sigma_{\rm GMC} = 2, 4, 8, 16 \kpc^{-2}$.}
\end{figure*}

A number of studies of disc heating via GMCs have been made, either via
numerical simulations or via more theoretical approaches, {\em e.g.} by
numerically integrating the Fokker-Planck equation.  Our aim here is to compare
our simulation results with previous work, and to compare with the
post-Hipparcos observations of the AVR.

Observations of the gas and dust in GMCs reveal a complex, fragmented structure
that is often described as self-similar or fractal.  We model our simulation
GMCs by spherical mass distributions of uniform density.  In many works
\cite{villu83,villu85} the GMCs have been simulated as having a softened point
potential (e.g. the Plummer model). However, when taking into account the
complex structure of GMCs we have preferred using an homogeneous density
distribution; we have thus modelled our GMCs as homogeneous spheres. The disc
heating is not very sensitive to these differences in the modelling.

Because most of the mass of interstellar matter is in the high end of the
molecular cloud mass function, it is sufficient to simulate only the most
massive GMCs, having typical masses of $M_{\rm GMC}=1 \times 10^6 \Msun$ and
diameters of $d_{\rm GMC}=100 \parc$ \cite{scosa86}. Because GMCs do not
interact with each other, the velocity dispersion of the GMC population is
frozen during the simulation run. The observed velocity dispersion of the
Galactic GMC population is so low that there must be some kind of physical
mechanism for cooling it. There is some evidence that infalling gas into the
Galactic disc cools down the interstellar gas. Whatever the physical mechanism,
we can emulate this situation simply by allowing no interactions between the
GMCs.

We start the simulations with both the tracer stars and the GMCs having a low
(cold) velocity dispersion, simulating that the stars have just been born from
the gas clouds, and share their kinematics.  The values used are taken from
Dehnen \& Binney \shortcite{debi}. The total velocity dispersion $\sigma$ is
decomposed into three components in the $x, y, z$ directions of
$(\sigma_x,\sigma_y,\sigma_z)$, with $\sigma_x=14$ \kms\ and $\sigma_z=6$ \kms\
($\sigma_y$ follows from the $\sigma_x$ and $\Omega / \kappa$-ratio). The
initial velocity distribution is Gaussian. The initial vertical structure is
Gaussian. Strictly speaking the vertical structure should obey a
$\sech^2$-distribution in order to be isothermal, but in practice the
$\sech^2$-distribution is very close to a Gaussian distribution in numerical
simulations with a limited number of particles.

In the GMC runs the simulation box size $2 \times 2 \times 4\kpc^3$ was found
to be large enough for the calculation of gravitational perturbations. After
testing it was found that an integration step size $\Delta t = T_{\mathrm
orb}/1000$ is sufficiently small for these runs, where $T_{\mathrm orb} \simeq
237.23 \times 10^6$ years. At the Solar radius the radial distribution of
molecular clouds is falling of rapidly with increasing radius. In the vicinity
of the Solar neighbourhood the average density of molecular clouds is
approximately $\Sigma_{\rm GMC} \simeq 5 \Msun / \pc^2 = 5 \times 10^6 \Msun /
\kpc^2$ \cite{scosa86}.

The evolution of the stellar velocity dispersion is plotted in Figure 3(a) for
four different assumed GMC number densities in the patch: $\Sigma_{\rm GMC}
=$2, 4, 8, 16 /$\kpc^2$.  As most of the mass is at the high-mass end of the
spectrum, the case $\Sigma_{\rm GMC} = 4 /\kpc^2$ is closest to the present
value for the Solar neighbourhood.

We have fit the simulations with an age-velocity relation (AVR) of the
form given in Eqn \ref{fit}. The second curve, from bottom to top, in
Figure 3(a) shows the fit with the most appropriate values for the
Solar neighbourhood.  The velocity dispersion is fitted only for the
values $\sigma_{tot} > 30 \kms$, because the stellar heating by GMCs
is known to obey a different power law when the stellar scale height
is less than cloud scale height \cite{lacey84}. Our fits are thus
valid when the stellar population has a larger scale height than the
GMC population. The velocity dispersion in the simulations is observed
to grow as $\sigma(t) \propto t^{0.21}$ (see Table 2). This is
slightly less than found in other studies ($\sigma(t) \propto
t^{0.25}$ using different methods; i.e. analytical treatment in Lacey
\shortcite{lacey84}, and ``scaled'' disc simulations in Villumsen
\shortcite{villu85}). The least-square-fit-values for $\alpha$ are
shown in Table \ref{gmctable} individually for the total
$\sigma_{tot}$, the radial $\sigma_x$ and the vertical component
$\sigma_z$ of the velocity dispersion. The vertical velocity
dispersion is observed to grow more efficiently than the velocity
dispersion in the horizontal plane.  This tendency is in accoradance
with the earlier results \cite{villu85}, but our values for the
$\alpha$ are somewhat smaller.

\setcounter{figure}{3}
\begin{figure*}
\label{fig2e}  
\input epsf
\center
\leavevmode
\epsfxsize=0.9
\columnwidth
\epsfbox{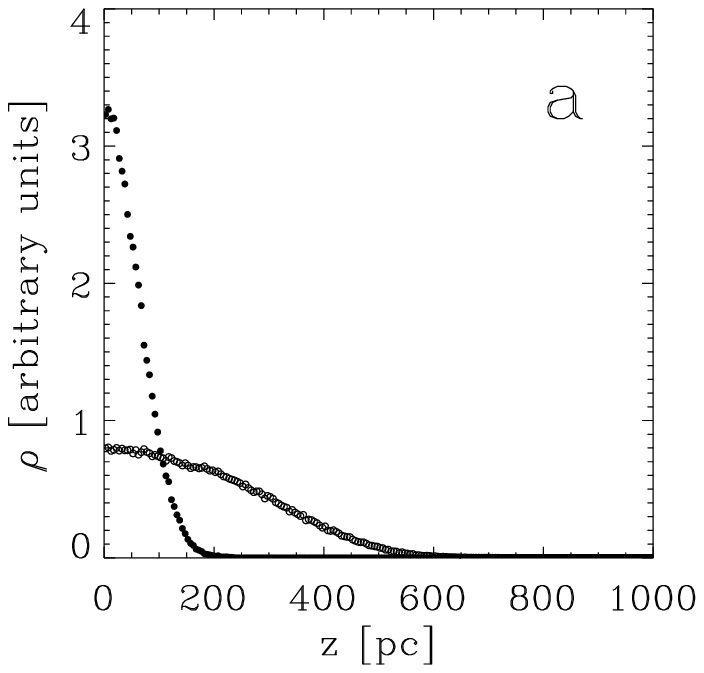}
\epsfbox{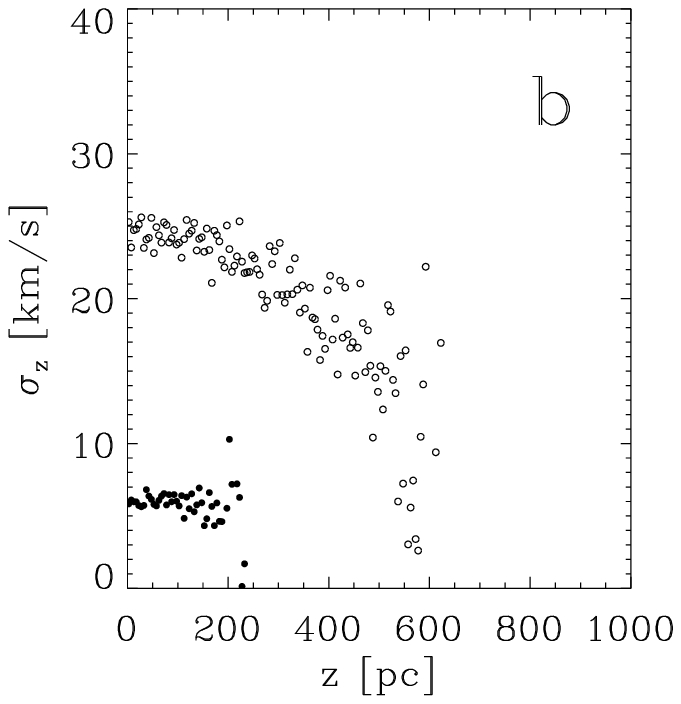}
\caption{Panel (a): the vertical density profile of the stellar disc at the
start (filled circles) and at the end (open circles) of the simulation for a
GMC number density of $\Sigma_{\rm GMC}=16 /\kpc^2$.  Panel (b): the vertical
component of the velocity dispersion as a function of vertical height at the
beginning (filled circles) and in the end (open circles) of the corresponding
simulation.}
\end{figure*}

The ratio $\sigma_z / \sigma_x$ of the vertical velocity dispersion to the
radial velocity dispersion is plotted in a Figure 3(b). The ratio is clearly a
function of the number density of the GMCs.  At the end of the simulation this
ratio varies between about 0.45 and 0.60; i.e. the vertical heating is
relatively most effective when the GMC number density is highest.  However, all
the values fall within the observational limits $\sigma_z / \sigma_x \simeq 0.5
\pm 0.1$. Coincidentally the mean value $\sigma_z / \sigma_x \simeq 0.5$ is
produced by the case $\Sigma_{\rm GMC} = 4 /\kpc^2$ which is approximately the
present GMC (of mass $M_{\rm GMC} = 1 \times 10^6 \Msun $) number density in
the Solar neighbourhood.

The ratio of the azimuthal component to the radial component depends on
$\Omega$ and $\kappa$ and is $\sigma_y / \sigma_x \simeq 0.7$ for the Solar
neighbourhood. This is also what is observed in the simulations.

Our results seem to be in poor accordance with Lacey's
\shortcite{lacey84} results; he predicts $\sigma_z / \sigma_x \simeq
0.8$ for the present values of the epicyclic constants. He has also
obtained $\sigma \propto t^{0.25}$ for all the velocity components
when the stellar scale height is larger than the GMC scale height.

Lacey generalized the work of Spitzer and Scwarzschild
\shortcite{spit51,spit53} (who calculated the evolution of the stellar velocity
distribution function by numerically integrating the Fokker-Planck equation)
and calculated the evolution of all three components of the velocity dispersion
by solving the first moments of the Fokker-Plack equation. He assumed that the
velocity distribution always remains a triaxial Gaussian. Most importantly, the
stellar orbits are assumed to be predominantly perturbed by many distant, weak
encounters so that the evolution of the distribution function could be
described by the diffusion equation.

Lacey uses first-order epicyclic theory for the stellar orbits (as we do here)
and assumes the molecular clouds to be long lived (as here). However, for
simplicity he adopts circular GMC orbits, ignores Galactic shear and assumes
that the interaction time between the stars and the GMCs is short.

A simple test is to vary the GMC size: a simulation run with $\Sigma_{\rm GMC}
= 4 / \kpc^2$ and $d_{\rm GMC} = 20 \parc$ (which is similar to Lacey's cloud
size) yielded $\alpha = 0.23$ for the total velocity dispersion. This goes in
the expected direction as smaller cloud size will increase the strength of the
gravitational perturbation in the close encounters. We also find that the
vertical component of the velocity dispersion is increased relative to the
radial one; we observed $\sigma_z / \sigma_x \simeq 0.57$ at the end of the
simulation. This partly explains the difference in the results. The remainder
we ascribe to the assumptions in the analytical method (i.e. circular GMC
orbits, no Galactic shear and short star-GMC interactions).

Numerical simulations by Villumsen \shortcite{villu85,villu83} have yielded
$\sigma_z / \sigma_x \simeq 0.6$. He obtained similar results for the growth of
the velocity dispersion in the plane as Lacey \shortcite{lacey84}. On the other
hand, for the vertical velocity dispersion Villumsen obtained $\sigma_z \propto
t^{0.31}$, which is significantly larger than the value we obtained ($\alpha =
0.26$).

Unfortunately, comparison of our results with Villumsen's work is not
straightforward because he analyses what one might term ``scaled-up''
simulations. Specifically, in his model, the exponential Galactic disc, with
scale length $1.75\kpc$, contains 4000 GMCs with total mass $M_{tot} = 4 \times
10^9 \Msun$ outside $3 \kpc$ radius, the individual GMCs having mass of $M_{\rm
GMC} = 1 \times 10^6 \Msun$. This model is then scaled so that it could be
simulated with fewer and heavier GMCs: the scaling being based on theoretical
expectations (e.g. Lacey 1984) that in the diffusion process the efficiency is
proportional to the total number of GMCs ($N_{\rm GMC}$) and to the square of
the mass ($M_{\rm GMC}^2$) of a single GMC, {\em i.e.} $N_{\rm GMC} \times
M_{\rm GMC}^2 = {\rm constant}$. In the actual simulations 200 or 400 GMCs with
mass $M_{\rm GMC} = 3.2 \times 10^6 \Msun$ have been used. The GMC size is the
same as ours ($d_{\rm GMC} = 100 \parc$), but a Plummer model has been used for
the GMC potential and the cut-off radius for the gravitational interaction has
been set at $2 \kpc$. The numerical scheme is rather different from ours, but
the equations of motion are integrated with a fourth order Runge-Kutta scheme
as in our simulations. Villumsen utilised an integration step size $\Delta t =
3 \times 10^6$ years, corresponding to $\Delta t = T_{orb}/74$ at the Solar
radius. In light of these many differences it is not surprising how difficult
it is to compare our simulations with Villumsen's.

One noticeable difference is, though, in the integration step size.
Villumsen \shortcite{villu85} reported that relative numerical error
in energy was less than $0.01 \%$ in an unperturbed
simulation. However, the errors in the numerical orbit integration
will be worst in close encounters between stars and massive GMCs.
Based on our experience with our numerical scheme, a considerably
larger step size than the one adopted by us induces artificial heating
in the system.

\setcounter{table}{1}
\begin{table}
\label{gmctable}
\caption{Fitting of the GMC heating. The least-square-fit of the exponent
$\alpha$ of the Eq. \ref{fit} is tabulated for the simulation runs along
with the number density of GMCs.}
\begin{tabular}[t]{l l l l}
\hline \\
$\Sigma$ & $\alpha$ for $\sigma_{tot}$ & $\alpha$ for
$\sigma_{x}$ & $\alpha$ for $\sigma_{z}$  \\ 
$[1/\kpc^2]$  &       &        &     \\
\hline \\
$\hspace{5pt} 2$ & $0.207 \pm 0.014$ & $0.198 \pm 0.019$  & $0.263 \pm 0.042$ \\
$\hspace{5pt} 4$ & $0.215 \pm 0.010$ & $0.208 \pm 0.015$  & $0.265 \pm 0.031$ \\
$\hspace{5pt} 8$ & $0.217 \pm 0.008$ & $0.212 \pm 0.012$  & $0.263 \pm 0.024$ \\
$16$          & $0.217 \pm 0.007$    & $0.214 \pm 0.012$  & $0.252 \pm 0.015$ \\
avg           & $0.21$               & $0.21$             & $0.26$            \\
\hline \\
\end{tabular}
\end{table}

The vertical density distributions of the tracer stars are plotted in Figure
4(a) both at the beginning and the end of the simulation run with $\Sigma_{\rm
GMC} = 16 /\kpc^2$. The initial density distribution (filled circles) is
exactly Gaussian with scale height of $z_{\circ} = 61 \parc$.  At the end of
the simulation run the vertical structure of the disc (open circles) is still
almost Gaussian with an obviously much larger scale height ($z_{\circ} \approx
230 \parc$). Compared to a Gaussian distribution the high end tail is missing
and the central density maximum is slightly suppressed, i.e. the profile is
more 'boxy', but still very close to a Gaussian.

The vertical velocity dispersion as a function of height is examined in Figure
4(b) for the same simulation. The initial velocity dispersion is isothermal
(filled circles). At the end of the simulation run, however, the velocity
dispersion is observed to decrease as distance from the horizontal plane
increases. Also numerical simulations by Jenkins \shortcite{je} have indicated
that in the coeval disc the molecular cloud heating mechanism does not retain
an isothermal velocity dispersion, but the velocity dispersion drops in higher
altitudes as in our results. His simulations also yield similar results
concerning the vertical density distribution: it is more 'boxy' than in the
isothermal disc.

To summarise the GMC section: our results rule out GMCs as the primary source
of disc heating, because the GMC number density required to heat the disc by
the correct amount is many times the observed GMC density in the Solar
neighbourhood (and our results indicate that GMC heating is even less effective
than earlier thought, {\em i.e.} $\sigma \propto t^{0.21}$).  However, contrary
to the earlier results, our simulations indicate that the velocity ellipsoid is
quite flat and consistent with the observational constraint on $\sigma_z /
\sigma_x$.

\section{Heating due to halo black holes}

\setcounter{figure}{4}
\begin{figure*}
\label{fig3a}   
\input epsf
\center
\leavevmode
\epsfxsize=0.9
\columnwidth
\epsfbox{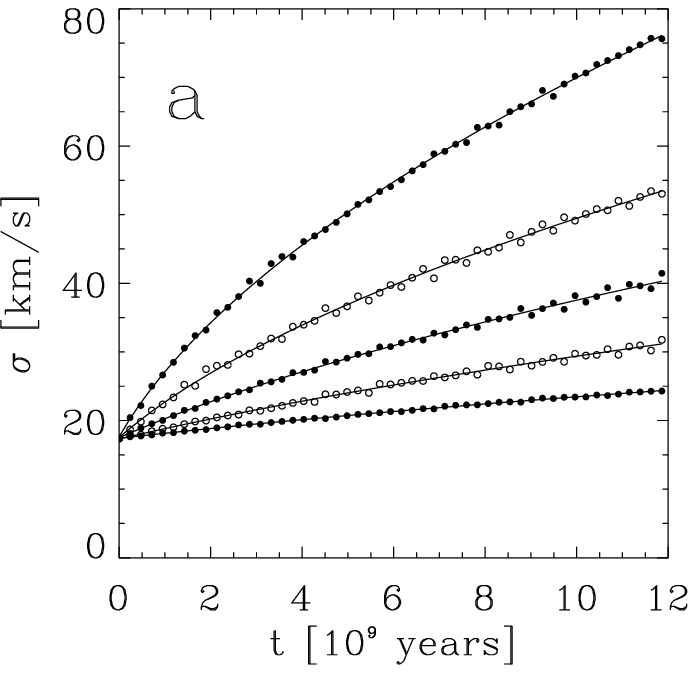}
\epsfbox{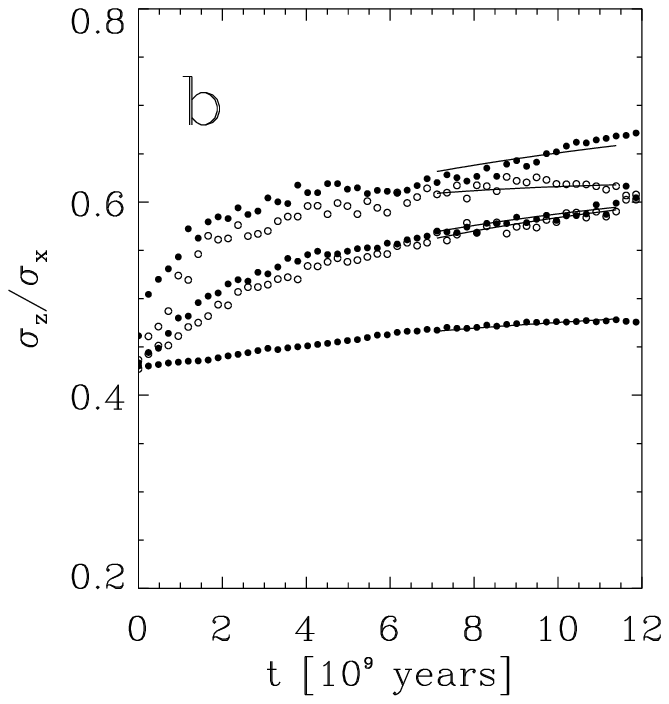}
\caption{Panel (a): the evolution of velocity dispersion $\sigma$ due to halo
black holes of mass $M = 1 \times 10^6 \Msun$. The curves represent simulation
runs with black hole number densities $\rho_{BH} = 4, 8, 16, 32, 64 /\kpc^3$
from bottom to top. Panel (b): the evolution of the ratio of the vertical
velocity dispersion to the radial velocity dispersion for the same
simulations.}
\end{figure*}

Another possibility for the heating of the Galactic disc would be due to
gravitational interaction with massive black holes in the Galactic dark halo,
which would make up part or all of the Galaxy's dark matter
\cite{lacey85}. These authors estimate that black holes of mass $M_{BH} \approx
2 \times 10^6 \Msun$ orbiting inside the Galactic halo could explain the
observed disc heating if they comprised the entire dark halo.

One should note that there are clear difficulties with the black hole
scenario. Based on estimates of the accretion rate of gas onto black holes
which are passing through the disc, \cite{mcdow} rule out more massive black
holes than $M_{BH} = 1000 \Msun$. Accretion should make such black holes
directly observable as optical or infrared objects, but no such objects have
been found. Lacey and Ostriker \shortcite{lacey85} estimate that the accretion
of the ISM onto the $M_{BH} \sim 10^6 \Msun$ halo black holes would create a
number of X-ray sources which should have been observed if they existed.  On
the other hand, recent work on the Galactic center black hole, which has a mass
of $2.6 \times 10^6$ \Msun \cite{genz98,nar98}, shows it to be a remarkably dim
X-ray source despite the high density of gas in which it sits, which indicates
that estimates of the X-ray flux of much nearer black holes of similar mass
passing through the disc ISM may be overestimated by several magnitudes.

Furthermore, this scenario has also purely dynamical difficulties. Very massive
black holes in the Galactic halo might also disrupt globular clusters in close
encounters within the Hubble time. Moore \shortcite{moore93} estimates that the
existence of the present population of low-mass globular clusters sets an upper
limit of $M_{BH} = 10^3 - 10^4 \Msun$ for the black hole masses. Klessen and
Burkert \shortcite{klebu96} estimate the upper limit to be $M_{BH} = 5 \times
10^4 \Msun$. However, there are also some difficulties when deducing these
upper limits. First of all, we do not know the initial mass function of
globular clusters. Secondly, as the evolution of a globular cluster is mainly
affected by BH encounters within its small core region and as these events are
rare, only few events during a Hubble time determine its fate. There will be thus
large statistical variations in globular cluster lifetimes.

Another limit on this scenario is introduced by the effects of dynamical
friction. Lacey and Ostriker \shortcite{lacey85} have estimated that black
holes of mass $M_{BH} = 2 \times 10^6 \Msun$ with initial orbital radii $r_{BH}
< 2 \kpc$ would spiral into the Galactic center within $t = 15 \times 10^9$
years. They assumed circular orbits for the black holes. The actual value for
the spiralling radius depends on the background density distribution which is
still not accurately known.

Given these uncertainties, we attempt in this paper to establish constraints
for the halo black hole population based purely on the kinematic evidence of
heating in the Galactic disc, and ignore these other physical constraints.

We test here the black hole scenario by simulating a patch of the local disc
embedded in a dark Galactic halo composed all or partly of massive black holes.
The black holes have been treated as point masses. We have simulated two
different mass objects: $M_{BH} = 1 \times 10^6 \Msun$ and $M_{BH} = 1 \times
10^7 \Msun$. We have choosen these masses because smaller masses have an
insignificant effect, and bigger masses are difficult to simulate in the
present method ({\em i.e.} using a patch --- the patch should be made so large
that one would need to begin to take into account global properties of the
Galaxy. In any case, black holes larger than $10^8 \Msun$ can be ruled out
because they would produce far too much heating, destroying the disc and/or
destroying halo globular clusters \cite{carr94}.

A softening of $\epsilon = 1 \parc$ has been used for the black holes in order
to avoid numerical problems.

The black holes are assumed to have zero net
rotation around the Galactic center and a 1-D Gaussian velocity dispersion of
$\sigma_x = \sigma_y = \sigma_z = 135 \kms$.  Black holes pass through the
simulation box with randomly chosen velocities drawn from these
kinematics. The stellar population has the same initial velocity dispersion as
in the GMC simulation runs.

\subsection{$\bf 10^6 M_{\sun}$ Black Holes}

In the runs with $M_{BH} = 1 \times 10^6 \Msun$, the patch size was set at $1
\times 1 \kpc^2$ in the horizontal plane. Different patch sizes were tested,
but this size patch was found to be adequate. We have used $2 \kpc$ for the
vertical extent of the simulation box.  The integration step size has to be
small enough in order not to produce artificial heating. After some testing we
found that $\Delta t = T_{orb}/8000$ is small enough.

We used 5000 stars in the patch and we investigated the effect of the number
density of black holes on the disc heating. Figure 5(a) shows the evolution of
the total velocity dispersion in five different cases: $\rho_{BH} = 4, 8, 16,
32, 64 / \kpc^3$. These number densities correspond to mass densities
$\rho_{BH} = 0.004, 0.008, 0.016, 0.032, 0.064 \Msun /\parc^3$, respectively.
For comparison, the mass density of the dark halo in the Solar neighbourhood is
$\rho_{H} \sim 0.01 \Msun /\parc^3$ \cite{holm00}, so the two higher number
density runs are clearly unrealistic, and their purpose is mainly to illustrate
the effectiveness or non-effectiveness of this heating mechanism.

It is already clear from Figure 5(a) that in order to account for the amount of
disc heating that is observed in the Solar neighbourhood there should be an
unrealistically large number of $10^6 \Msun$ black holes --- the total mass of
the black hole population would be larger than the mass of the whole Galactic
halo.

\setcounter{figure}{5}
\begin{figure*}
\label{fig3e}   
\input epsf
\center
\leavevmode
\epsfxsize=0.9
\columnwidth
\epsfbox{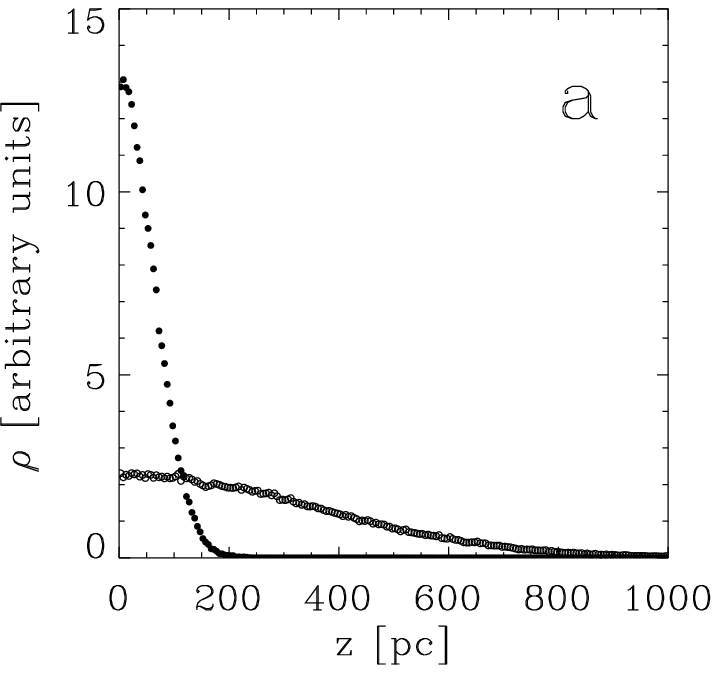}
\epsfbox{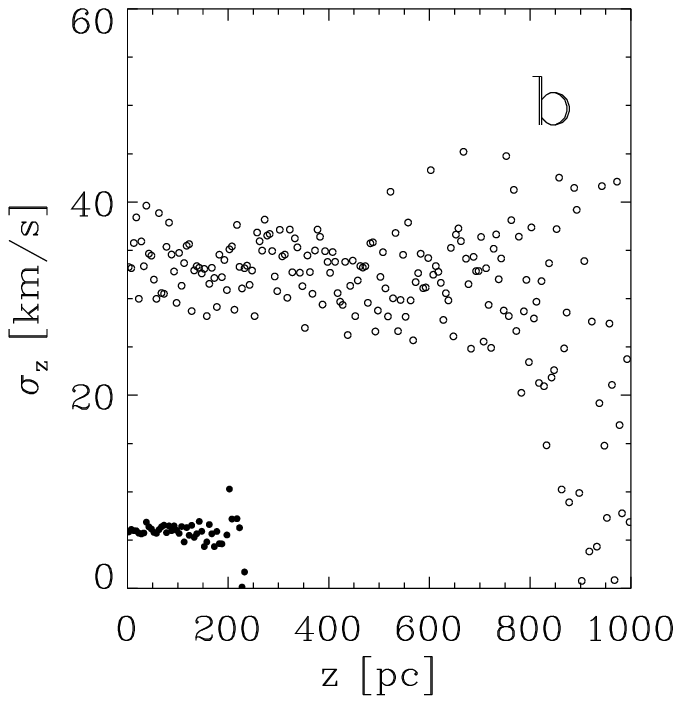}
\caption{Panel (a): The vertical density profile of the stellar disc is plotted
at the beginning and end of the simulation. Panel (b): the vertical component
of the velocity dispersion of the stellar disc is plotted as a function of
height at the beginning and end of the simulation. The adopted black hole mass
is $M_{BH} = 1 \times 10^6 \Msun$ and the number density is $\rho_{BH} = 32
/\kpc^3$.}
\end{figure*}

We have again used Eq. \ref{fit} in the least-square-fitting of the velocity
dispersion. The resulting exponent $\alpha$ for the total, radial, and vertical
velocity dispersions can be seen in Table 3. Our simulations verify the results
of Lacey \& Ostriker \shortcite{lacey85} that the total velocity dispersion is
proportional to $\sigma \propto t^{0.5}$ for the heating due to the halo black
holes.

They also found that the vertical-radial-axis-ratio is $\sigma_z / \sigma_x =
0.67$ in the non-selfgravitating case and $\sigma_z / \sigma_x = 0.55$ in the
self-gravitating case. For $10^6 \Msun$ black holes we find that $\sigma_z /
\sigma_x$ ranges from 0.47 to 0.67, the highest ratio being for the highest
black hole density (see Fig. 5b). The lowest value for $\sigma_z /
\sigma_x$-ratio is also closest to the observed value $\sigma_z / \sigma_x =
0.5 \pm 0.1$

The development of the vertical density profile in the simulation with $M_{BH}
= 1 \times 10^6 \Msun$ and $\rho = 32 / \kpc^3$ is shown in Figure 6(a). The
density profile remains very close to a Gaussian, and has a half width of
$z_\circ \approx 350 \parc$.  The vertical component of the velocity dispersion
is plotted in Figure 6(b) as a function of vertical distance from the
horizontal plane.  The velocities remain isothermal up to high altitudes, in
this case up to about 700 $\parc$ at the end of the simulation.

\setcounter{table}{2}
\begin{table}
\label{bh6table}
\caption{Fitting of the disc heating index $\alpha$ due to the black holes of
mass $M = 1 \times 10^6 \Msun$.}
\begin{tabular}[t]{l l l l}
\hline \\
$\rho$ & $\alpha$ for $\sigma_{tot}$ & $\alpha$ for
$\sigma_{x}$ & $\alpha$ for $\sigma_{z}$  \\ 
$[1/\kpc^3]$  &       &        &     \\
\hline \\
$\hspace{5pt} 4$ & $0.529 \pm 0.014$ & $0.507 \pm 0.018$& $0.597 \pm 0.030$ \\
$\hspace{5pt} 8$ & $0.514 \pm 0.026$ & $0.524 \pm 0.040$& $0.498 \pm 0.038$ \\
$            16$ & $0.535 \pm 0.057$ & $0.522 \pm 0.082$& $0.573 \pm 0.098$ \\
$32$             & $0.53 \pm 0.13$   & $0.52 \pm 0.19$  & $0.56 \pm 0.18$ \\
$64$             & $0.42 \pm 0.29$   & $0.42 \pm 0.40$  & $0.44 \pm 0.51$ \\
avg              & $0.51$            & $0.50$           & $0.53$          \\
\hline \\
\end{tabular}
\end{table}

\subsection {$\bf 10^7 M_{\sun}$ Black Holes}

\setcounter{figure}{6}
\begin{figure*}
\label{fig4a}   
\input epsf
\center
\leavevmode
\epsfxsize=0.9
\columnwidth
\epsfbox{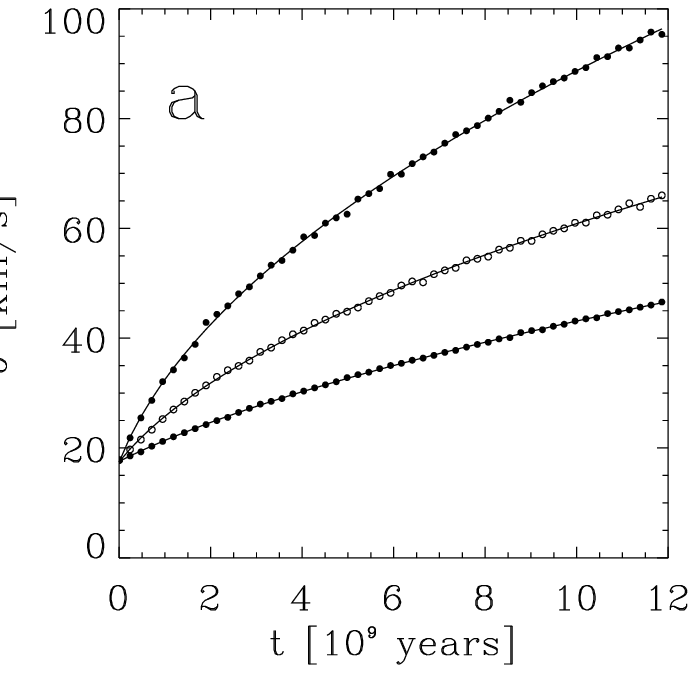}
\epsfbox{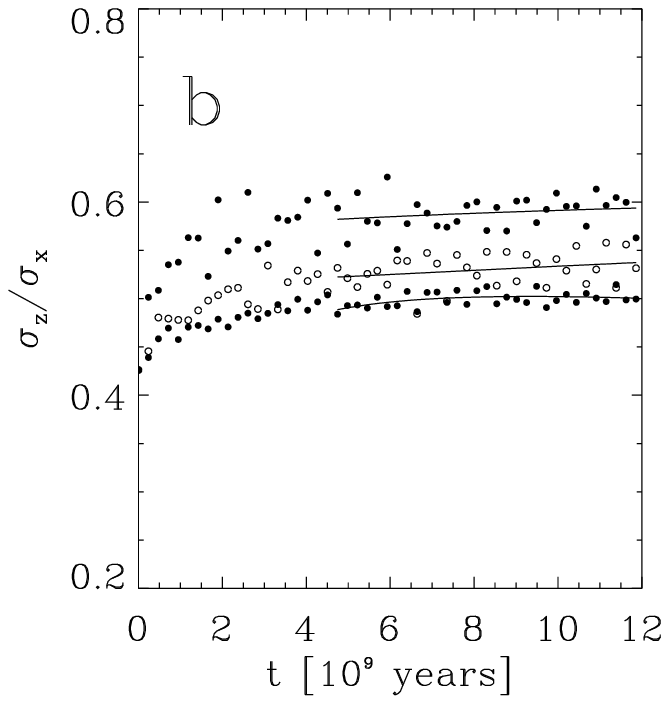}
\caption{Panel (a): the evolution of velocity dispersion $\sigma$ due to halo
black holes of mass $M = 1 \times 10^7 \Msun$. Panel (b): the evolution of the
ratios of the vertical velocity dispersion to the radial velocity dispersion is
plotted. The curves represent simulation runs with black hole number densities
$\rho_{BH} = 0.25, 0.5, 1.0 /\kpc^3$ from bottom to top.}
\end{figure*}

For the $M_{BH} = 1 \times 10^7 \Msun$ simulation runs a larger patch size was
necessary. The used patch size was $2 \times 2 \kpc^2$ in the horizontal plane
and $4 \kpc$ for the vertical size. The main reason for increasing the patch
size is that we have to calculate gravity from larger distances with the more
massive perturbers.

The cutoff radius for calculating gravity was thus $R_g = 1 \kpc$.  Testing
showed that a smaller integration step size was also needed, and we adopted
$\Delta t = T_{orb}/16000$.

For these simulations, 10,\,000 stars were placed in the patch. We used black
hole number densities of $\rho_{BH} = 0.25, 0.5, 1.0 /\kpc^3$, which correspond
to mass densities $\rho_{BH} = 0.0025, 0.005, 0.01 \Msun /\parc^3$,
respectively. The evolution of the total velocity dispersion is plotted in
Figure 7(a). It is interesting to note that it would take about 8 billion years
to heat up the stellar disc from $\sigma \simeq 18 \kms$ to $\sigma \simeq 80
\kms$ if the total mass of the halo consisted of $M_{BH} = 1 \times 10^7 \Msun$
mass black holes. There are some ambiguity in the observations: some results
indicate that the oldest stellar population would have a total velocity
dispersion only of $\sigma \simeq 50 \kms$. In this case the required amount of
heating would be produced if only half the halo mass were in the form of these
black holes. The RMS-fits of $\alpha$ for different components of the velocity
dispersion are found in Table 4.

In addition to the fact that these black holes broadly produce the right amount
of disc heating, they also distribute the heating in the three components
consistently with observations.  The $\sigma_z / \sigma_x$-ratios, plotted in a
Figure 7(b), range from 0.5 to 0.6, the highest black hole number density
producing highest $\sigma_z / \sigma_x$-ratio. Because of the ambiguities in
the observed AVR, the massive halo black holes can not be ruled out as being
responsible for the heating of the Galactic disc. However, if more weight is
put on the post-Hipparcos AVR determinations, then the value $\alpha=0.5$ could
be considered being too high.

Our results are in mild disagreement with the theoretical expectations
of Lacey and Ostriker \shortcite{lacey85}. They estimated that in
order to heat the stellar disc from $\sigma \approx 10 \kms$ up to
$\sigma \approx 80 \kms$ in $15$ Gyr the halo ($\rho = 0.01 \Msun
/\parc^3$) should be comprised of black holes of mass $M_{BH} = 2
\times 10^6 \Msun$. Our simulations show that even $M_{BH} = 1 \times
10^7 \Msun$ would not be enough to do the job. If the initial velocity
dispersion were around $\sigma \simeq 10 \kms$, the disc would be
heated only to $\sigma \simeq 60 \kms$ in $15$ Gyr. (Note that
Fig. 7(a) may be deceiving because a higher initial velocity
dispersion, $\sigma \simeq 18 \kms$, was adopted).

\setcounter{table}{3}
\begin{table}
\label{bh7table}
\caption{Fitting of the disc heating due to the black holes of mass
$M = 1 \times 10^7 \Msun$. Also the
standard deviation for the fitted parameter is given as an error estimate.}
\begin{tabular}[t]{l l l l}
\hline 
$\rho$        & $\alpha$ for $\sigma_{tot}$ & $\alpha$ for $\sigma_{x}$  &
$\alpha$ for $\sigma_{z}$  \\ 
$[1/\kpc^3]$  &   &  &     \\
\hline 
$  0.25$ & $0.52 \pm 0.04$   & $0.53  \pm 0.06$  &$0.50 \pm 0.04$   \\
$  0.5$  & $0.48 \pm 0.02$   & $0.48  \pm 0.02$  &$0.50 \pm 0.04$   \\
$  1.0$  & $0.503 \pm 0.008$ & $0.502 \pm 0.012$ &$0.509 \pm 0.016$ \\
avg      & $0.50$            & $0.50$            &  $0.50$          \\
\hline 
\end{tabular}
\end{table}

\section{Combined effect of GMCs and halo objects}

\setcounter{figure}{7}
\begin{figure*}
\label{fig5a}   
\input epsf
\center
\leavevmode
\epsfxsize=0.9
\columnwidth
\epsfbox{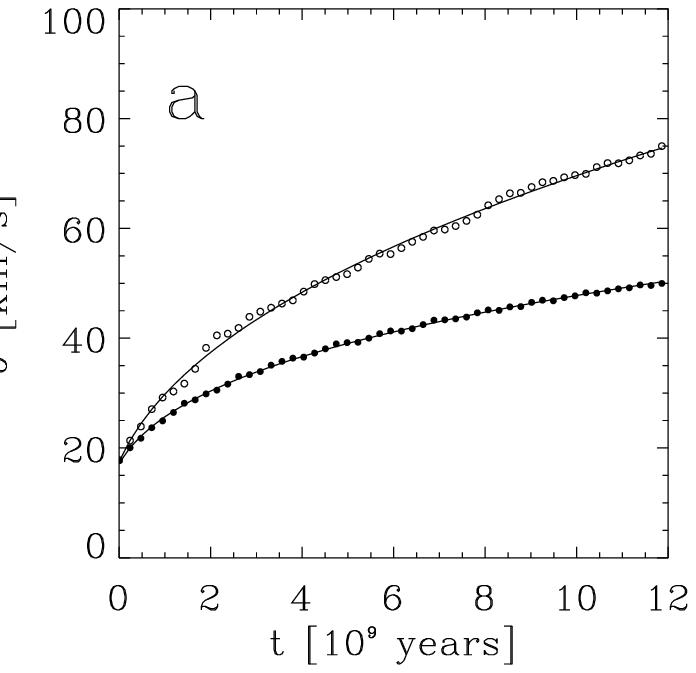}
\epsfbox{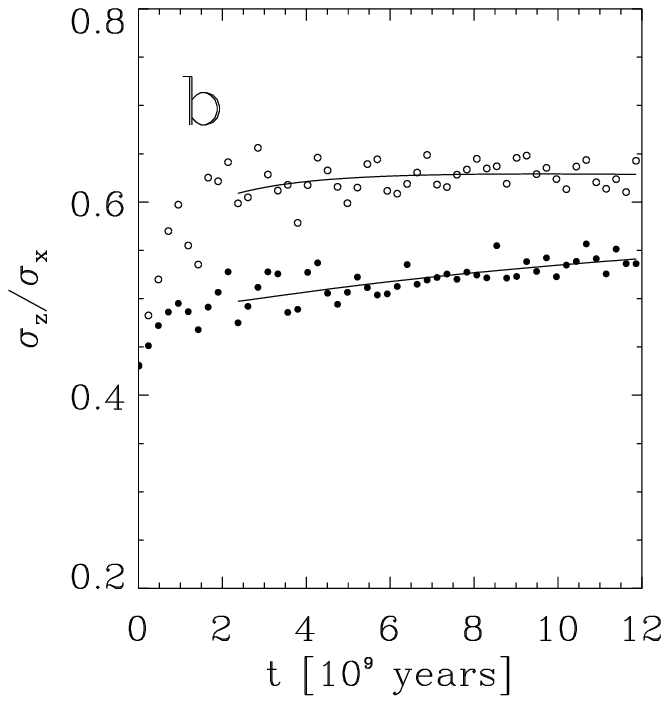}
\caption{Panel (a): the total velocity dispersion of the stellar disc. The
combined effect of the GMC and black hole perturbations are seen with two
different masses of the halo black holes: $M_{BH} = 1 \times 10^6 \Msun$
(filled circles) and $M_{BH} = 1 \times 10^7 \Msun$ (open circles). Panel (b):
the evolution of the ratios of the vertical velocity dispersion to the radial
velocity dispersion.}
\end{figure*}

In order to test the combined effects of the GMCs and halo black holes on the
disc heating we performed two further sets of simulations. We know
approximately the number density of the giant molecular clouds in the Solar
neighbourhood, but none of the parameters related to the possible black hole
population are directly known, except for the local density of the dark halo.

The simulations seen in Figure 8 differ only by the number density and mass of
the halo black holes. In the run of $M_{BH} = 1 \times 10^6 \Msun$ the number
density of the black holes is $\rho_{BH} = 8 /\kpc^3$ corresponding to
$\rho_{BH} = 0.008 \Msun /\pc^3$. In this scenario most of the halo would thus
consist of black holes. The combined effect of these black holes and GMCs would
be enough to heat the disc to $\sigma = 50 \kms$ in 12 Gyr (filled circles in
Figure 8(a)), which is somewhat less than what is actually observed (however,
see also Dehnen \& Binney \shortcite{debi} who report a maximum velcocity
dispersion $\sigma \simeq 50 \kms$ for the oldest stellar population). The
growth of the velocity dispersion is now intermediate between the pure black
hole and the GMC cases alone, $\alpha = 0.31$.

The ratio of the vertical and radial components of the velocity dispersion is
plotted in Figure 8(b). In this case (filled circles) the ratio is around
$\sigma_z / \sigma_x \simeq 0.53$ after $t \simeq 10$ Gyr. This value fits well
within observational error.

The open circles in Figure 8 show the simulation with halo black holes of mass
$M_{BH}= 1 \times 10^7 \Msun$, and a number density of $\rho_{BH} = 0.5
/\kpc^3$ which corresponds to $\rho_{BH} = 0.005 \Msun /\pc^3$.  We can
reproduce the disc heating very nicely with this setup, with the velocity
dispersion rising to $\sigma \simeq 75 \kms$ over 12 Gyr, very similar to the
observations. The fitted heating law is within the observational limits at
$\alpha = 0.42$. The ratio of $\sigma_z / \sigma_x$ is about 0.63, which is
also within the observational error limits.

It is interesting to note the high $\sigma_z / \sigma_x$ ratios that the
combined perturbations of BHs and GMCs produce. Why is the $\sigma_z /
\sigma_x$ ratio not somewhere between the values obtained from the simulations
in which BHs and GMCs act alone?  The answer lies in the fact that $\sigma_z /
\sigma_x$ ratio has a tendancy to grow when the number of perturbers is
increased, as seen in both the GMC and BH simulation sets.  When we increase
the total number of perturbers, even if some of them are BHs and some GMCs, the
$\sigma_z / \sigma_x$ ratio is bound to increase.

\section{Conclusions}

There is firm observational evidence for the heating of the stellar disc with
time.  We have assembled the available post-Hipparcos samples of ages and
kinematics for nearby stars and fit a heating law of the form $\sigma(t) =
\sigma_\circ (1 + {t}/{\tau})^\alpha$ to the data, where $\sigma$ is the
stellar velocity dispersion. The observational data limit $\alpha$ to values
between about $0.3$ and $0.6$, i.e. despite the recent advances in measuring
the kinematics and ages of nearby stars due to the availablity of Hipparcos
data, the age velocity relation in the nearby disc remains poorly
determined. Fortunately, the ratio of the vertical component of the velocity
dispersion to the radial component $\sigma_z / \sigma_x \simeq 0.5 \pm 0.1$ is
much better constrained by available data and can be used to test our
simulations.

We simulate disc heating by modelling a patch of the Galactic disc populated
with tracer stars. We examine the effect of Giant Molecular Clouds (GMCs)
travelling on disc orbits through the patch, and of massive black holes
travelling through on plunging orbits characteristic of the dark halo.

In our simulations of GMC heating we consistently find slightly lower values
for $\alpha$ than earlier studies have found, which are based on either
theoretical estimates of the diffusion rate of stellar orbits, or scaled
simulations of the whole Galactic disc (with a small number of GMCs).  We find
$\alpha = 0.21 \pm 0.02$ for the evolution of the total velocity dispersion and
$\alpha = 0.26 \pm 0.04$ for the evolution of the vertical component of the
velocity dispersion. The ratio $\sigma_z / \sigma_x$ of heating in the vertical
and horizontal directions in the disc is found to depend on the adopted GMC
density.

The observed heating can not be explained by GMCs, because even a four-fold GMC
number density compared to the present day value in the Solar neighbourhood,
$\Sigma \sim 5 \Msun / \parc^2$ \cite{scosa86}, does not heat up the disc
enough.  However, coincidentally we find that for the present day GMC number
density the ratio $\sigma_z / \sigma_x \simeq 0.5$ which is consistent within
the observational limits for stars in the local disc.

We have also run simulations of the local disc heating being caused by massive
($10^6$-$10^7$ \Msun) black holes of the dark halo penetrating the disc and
perturbing the stellar orbits.  Our simulations verify analytically obtained
results in the literature, that the heating index is of order $\alpha =
0.5$. Even if the entire halo consisted of $M_{BH} = 1 \times 10^6 \Msun$ black
holes, the resulting heating of the stellar disc would be much less than what
is observed. However, if the whole halo were comprised of black holes of order
$M_{BH} = 1 \times 10^7 \Msun$ the disc would be heated sufficiently. They
would heat the disc in such a way that the ratio $\sigma_z / \sigma_x$ would
increase from $0.5$ to $0.6$ which is still within the error limits of the
observations.

We have examined how the stellar disc would be heated by the combination of
GMCs and halo black holes. Adopting a GMC density consistent with the present
observed value and a dark halo comprised of $M_{BH} = 1 \times 10^6 \Msun$
black holes would only heat the disc up to $\sigma \simeq 50 \kms$ in 12 Gyr,
which is inconsistent with most of the observations. On the other hand, if half
of the halo were made of $M_{BH} = 1 \times 10^7 \Msun$ black holes, they could
with the GMCs heat the disc up to $\sigma \simeq 75 \kms$ in 12 Gyr, which is
consistent with observations, but the heating would also push the ratio of the
vertical velocity dispersion to the horizontal dispersion higher: $\sigma_z /
\sigma_x \simeq 0.63$, which is still barely within the observational
limits. In this case the heating index was observed to be $\alpha = 0.42$.

Taking into account the observational uncertainties in the stellar disc heating
we can produce the right amount of the heating by a proper combination of
massive halo black holes, $M_{BH} = 1 \times 10^7 \Msun$, and the right number
density of GMCs in the Solar neighbourhood. Also the ratio $\sigma_z /
\sigma_x$ and the heating index $\alpha$ fall within observational error
limits. We do not regard this as a particularly satisfactory solution, because
there are other potential causes of heating of stellar orbits, such as spiral
arms, which we plan to investigate further in the simulation method presented
here.

\section*{Acknowledgments}

We are grateful to Burkhard Fuchs, Johan Holmberg, Heikki Salo, Jesper
Sommer-Larsen, Alexandr Myll\"ari and Helio Rocha-Pinto for valuable
discussions. We thank the Academy of Finland for supporting this work, through
its support of the ANTARES space research program.

{} 

\end{document}